\documentclass[aps,prl,twocolumn,amsmath,amssymb,superscriptaddress,floatfix]{revtex4-1}

\usepackage{graphicx}
\usepackage{multirow}
\usepackage{color}
\usepackage{bm}
\usepackage{hyperref}

\renewcommand{\Vec}[1]{{\bm{#1}}}

\def\t#1{\textrm{#1}}
\def\ket#1{|#1\rangle }
\def\bra#1{\langle #1 |}
\def\braket#1{\langle #1 \rangle}
\def\n{\nonumber \\ }

\begin{document}
\title{Topological nature of nonlinear optical effects in solids
}
\author{Takahiro Morimoto}
\affiliation{Department of Physics,
University of California, Berkeley, CA 94720}
\author{Naoto Nagaosa}
\affiliation{RIKEN Center for Emergent Matter Science 
(CEMS), Wako, Saitama, 351-0198, Japan}
\affiliation{Department of Applied Physics, The University of 
Tokyo, Tokyo, 113-8656, Japan}

\date{\today}

\begin{abstract}
There are a variety of nonlinear optical effects including
higher harmonic generations, photovoltaic effects, and 
nonlinear Kerr rotations.
They are realized by
the strong light irradiation to materials that results in nonlinear 
polarizations in the electric field.
These are of great importance in studying the physics of 
excited states of the system as well as for applications to
optical devices and solar cells. Nonlinear properties of materials
are usually described by the nonlinear susceptibilities,
which have complex expressions including 
many matrix elements and energy denominators.   
On the other hand, a nonequilibrium steady state under a  
electric field periodic in time has a concise description
in terms of the Floquet bands of electrons
dressed by photons.
Here, we theoretically show by using the Floquet formalism 
that various nonlinear 
optical effects, such as the shift current in noncentrosymmetric 
materials, photovoltaic Hall response,
and photo-induced change of order parameters
under the continuous irradiation of monochromatic light, 
can be described in a unified fashion 
by topological quantities involving the Berry connection and Berry curvature.
It is found that vector fields defined with the Berry connections 
in the space of momentum and/or parameters 
govern the nonlinear responses. 
This topological view offers a new route to design the
nonlinear optical materials.  
\end{abstract}

\maketitle

\textit{Introduction ---}

Under strong light irradiation, materials show electric polarization
$\mathcal{P}$ or current $J= d\mathcal{P}/dt$ which are nonlinear functions of the 
electric field $E$.
These nonlinear optical responses (NLORs) form one of the most 
important research fields in condensed matter physics 
\cite{Bloembergen,Boyd},
since the nonlinearity often plays a crucial role in optical devices.
NLORs are also of crucial importance
for the solar cell action. The photo-current in a solar cell is usually described
by two processes, i.e., the generation of electron-hole pairs or 
excitons, and the separation of electrons and holes by the potential gradient
in the p-n junctions. 
A recent remarkable advance is the discovery of  
large efficiency of the solar cell action in perovskite oxides with noncentrosymmetric
crystal structure~\cite{Grinberg,Nie,Shi,deQuilettes,Bhatnagar}. 
One promising scheme that describes this phenomenon is the 
shift-current induced by the band structure 
without the inversion symmetry~\cite{Kraut,Young-Rappe,
Young-Zheng-Rappe,Cook}. 

While nonlinear optical processes described above involve high energy excited states,
the ground state and low energy excited states are 
sometimes characterized by the topological nature of the Bloch wavefunctions.
Specifically, the Berry connection and curvature of wave functions 
determine the ground state properties and the low energy transport 
phenomena. 
Such examples include
ferroelectricity~\cite{Resta}, 
quantum Hall effect~\cite{TKNN,Thouless}, 
anomalous Hall effect~\cite{AHE},
spin Hall effect~\cite{SHE1,SHE2}, 
and topological insulators~\cite{TI1,TI2,Ryu-review},
and ideal dc conduction~\cite{Hetenyi}.
Quantum mechanical wavefunctions can be regarded as 
geometrical objects in the Hilbert space 
because the inner product and distance are defined for them.
This is especially the case in solids, since the Bloch wavefunctions are
grouped into several bands separated by the energy gaps, and each 
band $n$ is regarded as a manifold in the Hilbert space. This manifold is   
characterized by a connection $\bm{a}_n (\bm{k})$ that relates neighboring two
wave functions in the crystal momentum ($k$-)space as
\begin{align}
\bm{a}_n (\bm{k}) = -i \bra{u_{n \bm{k}} } \nabla_{\bm{k}} u_{n \bm{k}} \rangle
\label{eq:vector}
\end{align}
where $\ket{ u_{n \bm{k}} }$ is the periodic part of the Bloch wave function.
One can also extend this concept to a generalized space including some
parameters $Q$'s characterizing the Hamiltonian such as the atomic 
displacement. Equation (\ref{eq:vector}) has the meaning of the
intracell coordinates~\cite{Adams} where the real-space
coordinate $\bm{x}_c$ of 
the wavepacket made from the Bloch wavefunctions near $\bm{k}$
is represented by 
\begin{align}
x^\mu_c = i \frac{\partial}{\partial k_\mu} - a^\mu_n (\bm{k}).
\label{eq:Adams}
\end{align}
The second term comes from the nontrivial connection of the 
manifold for the band $n$ and upgrades the usual derivative in 
$\bm{k}$ to the gauge covariant derivative which is physically observable. 
Although $a^\mu_n (\bm{k})$ is a gauge dependent
quantity (subject to a change of phases of wave functions),
this correction can be understood as a band-dependent shift of the electron 
position arising from different linear combinations of atomic orbitals in the unit cell~\cite{Adams}. 
It is noted here that the vector potential 
$a^\mu_n (\bm{k})$ is related to the real space position
because of the canonical conjugation relationship between
$\bm{x}$ and $\bm{k}$.
 
The quantum Hall effect is a famous example where geometry of wave functions plays a crucial role in the low-energy transport.
The Hall conductivity $\sigma_{xy}$ can be represented by the
integral of the Berry curvature 
$\mathcal{F}_n(\bm{k}) \equiv [\nabla_{\bm{k}} \times \bm{a}_n(\bm{k})]_z$ 
over the occupied states~\cite{TKNN}. 
In the case of insulator, the integral with respect to $\bm{k}$ over the 
first Brillouin zone is quantized and called the Chern number. 
This leads to the quantized $\sigma_{xy}$, i.e., 
(integer) quantum Hall effect. Replacing one of the momentum, e.g., $k_y$, by
some parameter $Q$ characterizing the Hamiltonian, $\sigma_{xy}$ turns
into the electric polarization induced by the change in $Q$~\cite{Resta}. 
A nonvanishing Chern number in $\bm{k}$-$Q$ space is tied
to the quantum pumping~\cite{Thouless}.

However, these topological characterizations have been limited to ground 
state properties or linear responses to the weak external stimuli of low frequency.
This is because NLORs involve higher energy excitations such as 
particle-hole pairs which drive the quantum state out of the ground state manifold.
Conventional descriptions of the nonlinear responses are given by nonlinear susceptibility tensors $\chi$'s whose independent components are specified by the crystal symmetry and time-reversal symmetry. 
Microscopically, $\chi$'s have complex expressions including many matrix elements of the dipole moment along with energy denominators. 
These expressions usually do not give much information except for the trivial fact that the nonlinear responses show a resonance effect when the energy of light is nearly equal to the energy difference between the two states connected by the matrix elements.
The topological nature of responses to the strong and/or high frequency stimuli 
have not been explored thus far except for a few cases.

The shift-current is one of such a few nonlinear phenomena
whose geometrical meaning has been studied.
The photo-current is the current induced by light irradiation as is well known.
The induced photo-current $J$ is usually proportional to $E^3$
when the system preserves the inversion symmetry.
However, when the system lacks the inversion symmetry,
the photo-current $J$ can be proportional to 
$E^2$ and it is called ``shift-current''.
von Baltz and Kraut~\cite{Kraut} have derived a formula for this shift-current,
and related it to the intracell coordinates mentioned above. 
Specifically, it is 
expressed in terms of the phase $\varphi_{ij}(\bm{k})$ of the velocity matrix 
element $v_{ij}(\bm{k}) $ between the valence and conduction bands, and  
the Berry connection $\bm{a}_n (\bm{k})$ as
\begin{align}
\bm{J} \propto & E^2 \int d \bm{k} 
\delta( \epsilon_1(\bm{k}) - \epsilon_2(\bm{k}) + \hbar \omega)
\n
&\times |v_{12}(\bm{k})|^2 
\left[ \nabla_{\bm{k}} \varphi_{12}(\bm{k}) + \bm{a}_1 (\bm{k})
- \bm{a}_2 (\bm{k}) \right],
\label{eq:shiftcurrent}
\end{align}
where the subscripts 1 and 2 refer to the valence band and the conduction band, respectively,
$\epsilon_i(\bm{k})$ is the energy of the band, and 
$ \hbar \omega $ is the energy of the incident light.  
Note that this expression is gauge invariant due to the combination of 
$\nabla_{\bm{k}} \varphi_{12}(\bm{k})$ and $\bm{a}_{1,2} (\bm{k})$,
and it is remarkable in a sense that the vector potential itself appears
in the physical quantities. It is considered as a candidate mechanism of the high efficiency 
photovoltaic current in the solar cell action without the p-n junction
~\cite{Grinberg,Nie,Shi,deQuilettes,Young-Rappe,Young-Zheng-Rappe,Bhatnagar,Cook}.
We note that the photovoltaic Hall effect of two-dimensional Dirac fermions, e.g., in graphene,
has also been studied as a topological 
phenomenon~\cite{Oka,Kitagawa,Sentef15,McIver12,Jotzu}, where the circularly 
polarized light induces the Hall conductance 
$\sigma_{xy}$ proportional to $E^2$. 
In this case, the current $J$ is the third order effect, i.e., 
$\propto E^3$.
It is shown that the light induced $\sigma_{xy}$ is expressed by a similar formula to that
for the linear response; the only modification in the expression for $\sigma_{xy}$ is that
$\mathcal{F}_n(\bm{k})$ and the 
Fermi distribution function are replaced by those of the nonequilibrium Floquet bands.
 
In this paper, we study the topological nature of the nonlinear optical responses
by employing the Floquet two band models.
This formalism offers a general description of nonlinear responses when the following conditions are met: (i) only one frequency $\Omega$ is involved (the monochromatic light), (ii) mostly two bands are involved in the optical transitions, and (iii) a steady state is achieved.
We show that nonlinear optical responses of the even order of the external electric field $E$, such as photovoltaic effects and second harmonic generations, have geometrical meaning 
and are characterized by the Berry connection
in a generalized space including both the momentum $\bm{k}$ 
and parameters $Q$'s. 
In particular, we point out that these topological description is applicable for general noncentrosymmetric crystals that support the even order nonlinear responses. 
We also discuss that nonlinear dc Hall responses, which are nonlinear responses in the odd order of $E$ in general, 
are related to the Berry curvature of Floquet bands.
(It is noted however, that the topological description is limited to the dc output in this case of odd order responses.)
Moreover, we classify the nonlinear processes 
according to the presence or absence of the inversion ($P$) and time-reversal
($T$) symmetries in terms of the Berry connection and the Berry curvature. 
In order to demonstrate our general discussions,
we apply our formalism to a 1D model with inversion symmetry breaking which is a simple model of ferroelectric materials.
By doing so, we clarify the topological nature of a few nonlinear responses and the symmetry constraints to the nonlinear responses in an explicit way.

\begin{figure}[tb]
\begin{center}
\includegraphics[width=0.6\linewidth]{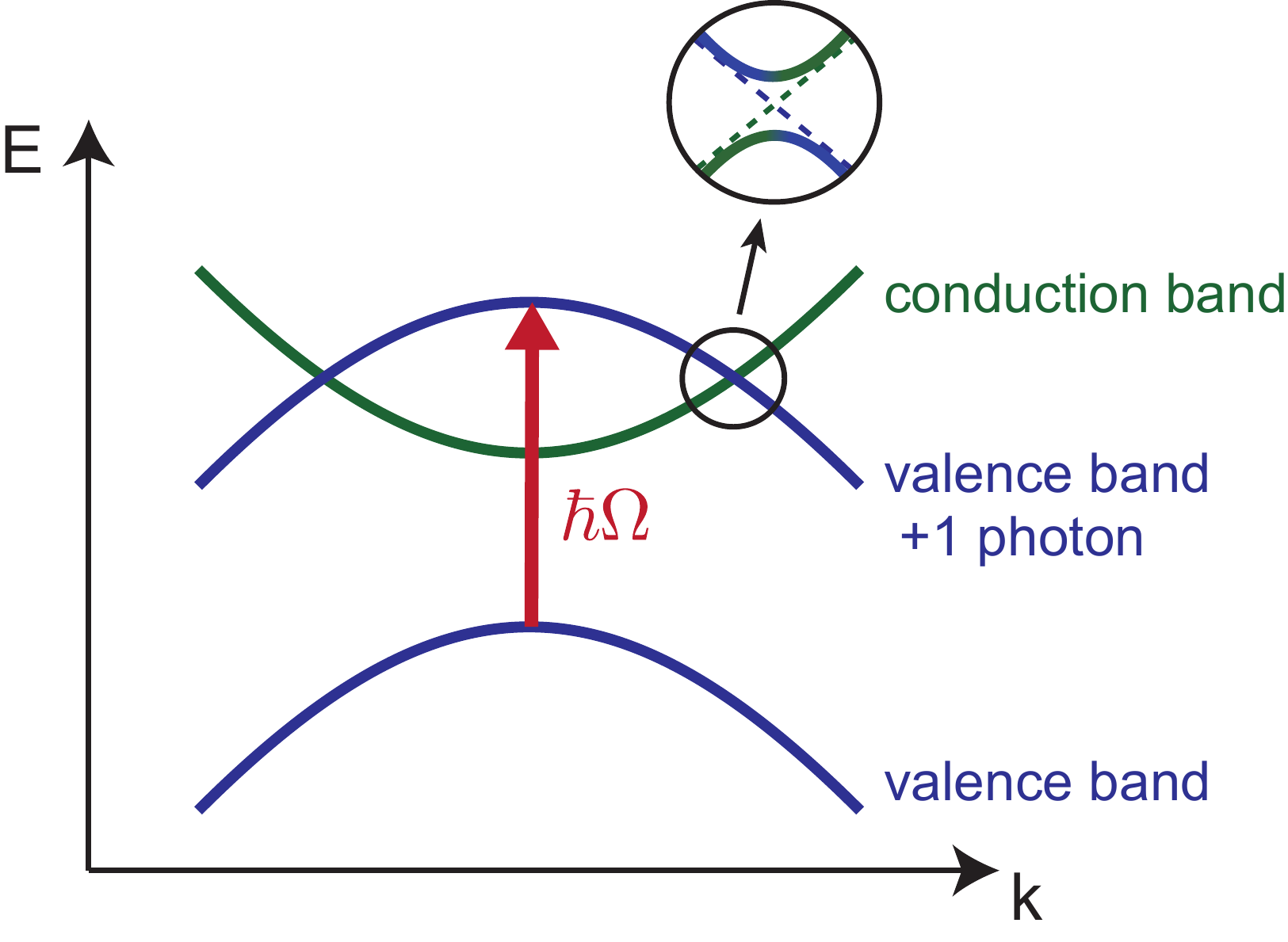}
\end{center}
\caption{
Schematic picture of the Floquet two band model.
Under the drive by the monochromatic light, energy bands evolve into Floquet bands which describe Bloch states dressed with photons.
When two Floquet bands cross, they show an anticrossing.
The nonequilibrium steady state (and hence, nonlinear optical responses) can be captured by studying this anticrossing of two Floquet bands.}
\label{fig: floquet anticrossing}
\end{figure}

\ \par
\ \par
\textbf{Results:}
\ \par

\textit{Floquet two band model ---}
We study nonlinear current responses 
by employing the Keldysh Green's function method combined with the Floquet 
formalism~\cite{Kohler,Jauho,Johnsen,Kamenev}.
(See Materials and Methods for details of the formalism.)
We focus on the two bands involved in the transition induced by the monochromatic light with the electric field $E(t)=E e^{-i\Omega t}+E^* e^{i\Omega t}$.
By using the Floquet bands, one can describe the nonequilibrium steady state as 
an anticrossing of a valence band dressed with one photon and 
a conduction band dressed with no photon,
which is schematically shown in Fig.~\ref{fig: floquet anticrossing}.
The anticrossing of these two Floquet bands 
is captured by the 
following Hamiltonian (the convention $\hbar=1, e=1$ is used hereafter): 
\begin{align}
H_F&=
\begin{pmatrix}
\epsilon^0_1 +\Omega & -i A^* v^0_{12} \\
iA v^0_{21} & \epsilon^0_2
\end{pmatrix}
\equiv
\epsilon+\Vec d \cdot \Vec \sigma,
\label{eq: two band model}
\end{align}
where
the subscripts 1 and 2 refer to the valence band and conduction band, respectively,
$\epsilon^0$ 
is the original energy dispersion (for $E=0$),
$A=E/\Omega$ and
$v^0=\partial H_0(A=0)/\partial k$.
The dc current operator is given by
\begin{align}
\tilde{v}&=
\frac{\partial H_F}{\partial k}
\begin{pmatrix}
v^0_{11} & -i A^* \left(\frac{\partial v^0}{\partial k}\right)_{12} \\
i A \left(\frac{\partial v^0}{\partial k}\right)_{21} & v^0_{22}
\end{pmatrix}
\equiv
b_0+\Vec b \cdot \Vec \sigma.
\end{align}
Physical quantities are obtained from the lesser Green's function 
which is given for the two band model as
\begin{align}
G^<&=
\frac{(\omega-\epsilon+i\Gamma/2+\Vec d \cdot \Vec \sigma)
\Sigma^<
(\omega-\epsilon-i\Gamma/2+\Vec d \cdot \Vec \sigma)}
{[(\omega-\epsilon+i\Gamma/2)^2-d^2][(\omega-\epsilon-i\Gamma/2)^2-d^2]}.
\end{align}
Here the lesser self-energy is given by 
$\Sigma^<=i\Gamma(1+\sigma_z)/2$.
This form of $\Sigma^<$ 
assumes that the system couples to a heat bath 
which has a uniform energy spectrum and the Fermi energy lying within the the energy gap of the system.
Thus we obtain the dc current expectation value as
\begin{align}
J&=-i\t{Tr}(\tilde{v} G^<) 
=\int d\bm{k} (j_1+j_2+j_3),
\label{eq: J three terms}
\end{align}
with
\begin{align}
j_1&=
\frac{\frac{\Gamma}{2}(-d_x b_y+d_y b_x)}{d^2+\frac{\Gamma^2}{4}},
\\
j_2&=
\frac{(d_x b_x+d_y b_y)d_z}{d^2+\frac{\Gamma^2}{4}},
\\
j_3&=
\frac{(d_z^2+\frac{\Gamma^2}{4}) b_z}{d^2+\frac{\Gamma^2}{4}}+b_0,
\end{align}
where $\t{Tr}$ denotes an integration over $\omega$ and $k$ and a trace for two by two matrices. 

The first term $j_1$ in Eq.~(\ref{eq: J three terms}) can be written with the 
Berry phase as follows.
First we write the denominator as
\begin{align}
-d_x b_y+d_y b_x&=
\t{Im}[(d_x +i d_y)(b_x-i b_y)] \n
&=
|A|^2 \t{Im}\left[v^0_{21} \left(\frac{\partial v^0}{\partial k} \right)_{12} \right]
,
\end{align}
with the current operator $v^0$ for the system without driving by $E$.
In the two band model, the matrix element of $\frac{\partial v^0}{\partial k}$
is written as
\begin{align}
\left(\frac{\partial v^0}{\partial k} \right)_{12}
&=\frac{\partial v^0_{12}}{\partial k}
-\bra{\partial_k u_1} v^0 \ket{u_2}
-\bra{u_1}v^0 \ket{\partial_k u_2} \n
&=v^0_{12} \left[
\frac{\partial}{\partial k}\log v^0_{12}
+ i a_{1}- i a_{2} +\frac{v^0_{11}- v^0_{22}}{\epsilon_1-\epsilon_2}
\right],
\label{eq: dv/dk}
\end{align}
with $a_{i}=-i \bra{u_i} \partial_k u_i  \rangle$.
Here, we used the identity 
$\ket{u_1}\bra{u_1}+\ket{u_2}\bra{u_2}=1$
for the two band model,
and $\bra{u_1} \partial_k u_2  \rangle = -v_{12}^0 /(\epsilon_1-\epsilon_2)$.
Thus we obtains
\begin{align}
d_x b_y-d_y b_x&=
|A|^2 |v^0_{12}|^2 R_k
\\
R_k &=
\left[
\frac{\partial \varphi_{12}}{\partial k}
+ a_{1}- a_{2}
\right],
\label{eq:Rk}
\end{align}
with $\varphi_{12}=\t{Im}(\log v^0_{12})$,
since $a_{i}$ and $v^0_{ii}$ are real.
The vector $R_k$ is called a shift vector,
which measures the difference of intracell coordinate between two band involved in the resonance.
We note that $R_k$ is a gauge invariant quantity,
where the Berry connection accompanies the $k$-derivative of the velocity operator to compensate the nontrivial parallel translation for the Bloch wave functions in $k$.
Then the contribution to the current expectation value is written as
\begin{align}
j_1 
&=
|A|^2 \frac{\frac{\Gamma}{2}}{d_z^2+|A|^2 |v^0_{12}|^2 +\frac{\Gamma^2}{4}}
|v^0_{12}|^2 
R_k. 
\n
&\cong  \frac{\pi |E|^2}{\Omega^2} 
\frac{\frac{\Gamma}{2}}{\sqrt{ \frac{|E|^2 |v_{12}^0|^2}{\Omega^2} +\frac{\Gamma^2}{4}}} \delta(d_z) 
|v^0_{12}|^2 
R_k,
\label{eq: j1}
\end{align}
where we have assumed in the second line 
that $\Gamma$ and $|A v_{12}^0|$ are much smaller than the
energy dispersion. 
If we further assume sufficiently small electric fields ($|A v_{12}^0| \ll \Gamma$),
this reduces to 
$j_1 \cong \frac{\pi |E|^2}{\Omega^2} 
\delta(d_z) 
|v^0_{12}|^2 
R_k$.
The second term $j_2$ in Eq.~({\ref{eq: J three terms}}) is rewritten by using 
$d_x b_x+d_y b_y=|A|^2 \t{Re}[v^0_{21} (\frac{\partial v^0}{\partial k} )_{12} ]$
as
\begin{align}
j_2 =
\frac{|A|^2 \t{Re}[v^0_{21} (\frac{\partial v^0}{\partial k} )_{12} ] d_z}{d^2+\frac{\Gamma^2}{4}}.
\end{align}
This contribution vanishes after integration over $k$ in the presence of the time-reversal symmetry (TRS),
because $j_2$ is odd under the TRS.
This can be understood from Eq.~(\ref{eq: dv/dk}) and the fact that $v_{ii}^0$ is odd under the TRS.
In a similar way, the contribution of $j_3$ in Eq.~({\ref{eq: J three terms}}) vanishes because $j_3$ is odd under the TRS.

To summarize, the photocurrent in the second order of $E$
 is given by
\begin{align}
J=
\frac{\pi |E|^2}{\Omega^2} 
\int d\bm{k} \delta(\epsilon_1^0-\epsilon_2^0 +\hbar \omega) 
|v^0_{12}|^2 
\left[
\partial_k \varphi_{12}
+ a_{1}- a_{2}
\right],
\label{eq: shift current formula}
\end{align}
in the presence of the time-reversal symmetry,
which reproduces the expression for the shift current~\cite{Kraut,Sipe,Young-Rappe}.
While we have focused on the two band model,
the result in Eq.~(\ref{eq: shift current formula}) can be extended to general cases involving more energy bands
by summing up contributions from any two bands satisfying the resonance condition.
We note that higher order correction to the above formula is captured by the factor
$\frac{\Gamma}{2}/\sqrt{\frac{|E|^2 |v_{12}^0|^2}{\Omega^2}+\frac{\Gamma^2}{4}}$
in Eq.~(\ref{eq: j1}).
This leads to a crossover of photocurrent from 
$J\propto E^2$ to $J\propto \Gamma E$ by increasing the intensity of the monochromatic light
and describes the effect of saturation of excitations.

\textit{Second harmonic generation ---}
The SHG is the nonlinear current response with the frequency $2\Omega$ induced by a monochromatic light 
$E(t)=E e^{-i\Omega t}+E^* e^{i\Omega t}$.
We show that the SHG is also described with geometrical quantity 
(i.e., Berry connections) in a similar manner to the shift current.
Here we consider the interband contribution to the SHG that involves two energy bands in the optical transition.
In this case, we can apply our approach based on the Floquet two band model.
Specifically, the SHG is contributed by two types of optical processes and accordingly two Floquet two band models:
$H_F$ in Eq.~(\ref{eq: two band model}) and
\begin{align}
H_F'&=
\begin{pmatrix}
\epsilon^0_1 + 2\Omega & -\frac{1}{2} (A^*)^2 \partial_k v^0_{12} \\
-\frac{1}{2} A^2 \partial_k  v^0_{21} & \epsilon^0_2
\end{pmatrix}
\equiv d_0' + \bm{d'}\cdot \bm{\sigma}.
\end{align}
The Floquet formalism also offers a concise description of time-dependent current responses, which is given in Eq.~(\ref{eq: J(t)}) in the Method section.
According to Eq.~(\ref{eq: J(t)}),
a contribution to  $J(2\Omega)$, which is the Fourier component of the current proportional to $e^{-2i\Omega t}$,
is written as
$-i v_{12}' (G^<)_{21}$ for each Floquet two band model,
where $v'$ are chosen so as to give the time-dependence of $e^{-2i\Omega t}$.
This is achieved by
$v_{12}'=iA(\partial_k v^0)_{12}$ in the case of $H_F$ and 
$v_{12}'=v^0_{12}$ in the case of $H_F'$.
By using Eq.~(\ref{eq: G< 21}) for $(G^<)_{21}$,
the interband contribution to the SHG is written as
\begin{align}
J(2\Omega)&\cong
i\frac{E^2}{2\Omega^2} \int d\bm k |v^0_{12}|^2 (\partial_k \varphi_{12}+a_1-a_2) 
\n
&\qquad \times
\left(
-\frac{1}{d_z+\frac{i\Gamma}{2}}
+
\frac{1}{2(d'_z+\frac{i\Gamma}{2})}
\right),
\end{align}
where we only kept nonvanishing terms in the presence of the TRS.
In particular, if we focus on the contribution to the SHG by the interband resonance
that involves a $\delta$-function with respect to the energy difference, 
such contribution is given by
\begin{align}
J(2\Omega)&\cong
\frac{\pi E^2}{2\Omega^2} \int d\bm k |v^0_{12}|^2 R_k 
\n
&\qquad \times
\left[
-\delta(\epsilon^0_1-\epsilon^0_2+\Omega)
+\frac{1}{2}\delta(\epsilon^0_1-\epsilon^0_2+2\Omega)
\right].
\end{align}
This indicates that the interband contribution to the SHG is characterized by the shift vector $R_k$ which is defined with Berry connections. 
Thus the SHG is generated by dynamics of excited electron-hole pair that experiences a shift of intracell coordinates in the transition between the valence and conduction bands and is naturally related to the Berry connection of the Bloch electron.

\textit{Third order nonlinear response ---}
Now we proceed to the third order nonlinear responses which are described by
\begin{align}
J_i=\chi^3_{ij} E(\omega)E(-\omega)E_j(\omega=0).
\end{align}
Here, the current $J_i$ is induced by 
the static electric field $E_j$ 
in the presence of pump laser light of the frequency $\omega$.

We focus on the nonlinear Hall response in the 
two-dimensional systems.
The nonlinear Hall response is obtained by
applying the linear response theory for the nonequilibrium steady states,
\begin{align}
\sigma_{xy}&=
\int_\t{BZ} d^2 \bm{k} 
\sum_i f_i (\Vec{\nabla} \times \widetilde{\Vec{a_i}})_z ,
\end{align}
and expanding it in the second power of $|E(\omega)|$~\cite{Oka}.
Here, the Berry connection $\widetilde{\Vec a}$ is defined for the Floquet states
that describes the nonequilibrium steady states,
and $f_i$ is the occupation of the $i$th Floquet state.
The wave functions of the Floquet two band model
in Eq.~(\ref{eq: two band model}) is given by
\begin{align}
u_1&=
\begin{pmatrix}
\cos \frac \theta 2 \\
\sin \frac \theta 2 e^{i\phi} \\
\end{pmatrix},
&
u_2&=
\begin{pmatrix}
-\sin \frac \theta 2 \\
\cos \frac \theta 2 e^{i\phi} \\
\end{pmatrix},
\end{align}
with
\begin{align}
\cos \theta& = \frac{d_z}{d},
&
\phi & = \tan^{-1} \left(\frac{d_y}{d_x}\right)=\varphi_{21}.
\end{align}
Then the Berry connections for Floquet bands $u_1$ and $u_2$ are given by
\begin{align}
\widetilde{\bm{ a_1}}&=\frac{1}{2}(1-\cos \theta)\Vec{\nabla} \phi 
+ \cos^2 \frac \theta 2 \bm{a_1} + \sin^2 \frac \theta 2 \bm{a_2}, \\
\widetilde{\bm{ a_2}}&=\frac{1}{2}(1+\cos \theta)\Vec{\nabla} \phi 
+ \sin^2 \frac \theta 2 \bm{a_1} + \cos^2 \frac \theta 2 \bm{a_2}, 
\end{align}
where $\bm{a_1}$ and $\bm{a_2}$ are Berry connections for the original bands with $E=0$.
The occupations are 
$f_1=(1+\cos\theta)/2$, and 
$f_2=(1-\cos\theta)/2$, 
since $f_i=(-i\Sigma^<_{ii})/\Gamma$ with
$-i\Sigma^</\Gamma =(\sigma_z+1)/2$.
The original Hall conductivity $\sigma_{xy}^0$ for $E=0$ is given by setting $\theta=0$.
Then one can obtain the photo-induced part of the Hall conductivity
$\sigma_{xy}^p \equiv \sigma_{xy}-\sigma_{xy}^0$
as
\begin{align}
\sigma_{xy}^p
&=
\frac 1 4 \int d^2 \bm{k} \sin^2 \theta [\Vec{\nabla} \times 
(\Vec{\nabla} \phi+ \bm{a_2}-\bm{a_1})]_z \n
&=\int d^2 \bm{k} \frac{A^2 |v^0_{12}|^2}{4(d_z^2+A^2 |v^0_{12}|^2)} \mathcal{F},
\label{eq: nonlinear Hall response}
\end{align}
where $\mathcal{F}\equiv (\Vec{\nabla} \times \bm{a_2}-\Vec{\nabla} \times \bm{a_1})_z$.
If we assume that $A$ is sufficiently small,
the Hall conductivity is given by
\begin{align}
\sigma_{xy}^p&=
\int d^2 \bm{k} \frac{\pi E}{4 \Omega} |v^0_{12}| \delta(d_z) \mathcal{F}.
\label{eq:sigmaxy1}
\end{align}
We can include the effect of relaxation by replacing the denominator 
in Eq.~(\ref{eq: nonlinear Hall response})
with $4(d_z^2+A^2 |v^0_{12}|^2+\Gamma^2)$,
which leads to 
\begin{align}
\sigma_{xy}^p&=
\int d^2 \bm{k}
\frac{\pi E^2}{4\Gamma \Omega^2} |v^0_{12}|^2 \delta(d_z) \mathcal{F}.
\label{eq:sigmaxy2}
\end{align}
This photo-induced 
Hall conductivity is proportional to $E^2$ and describes the third order nonlinear response.
It is also proportional 
to the relaxation time $1/\Gamma$ indicating that the nonlinear modulation 
arises from excited free electrons.
Equation (\ref{eq:sigmaxy1}) corresponds to the case of 
$A |v^0_{12}| \gg \Gamma $, while Eq. (\ref{eq:sigmaxy2}) to  the
case of $A |v^0_{12}| \ll \Gamma $.  Therefore these two equations 
describe the crossover from $\sigma_{xy}^p \propto E$ to $\propto E^2/\Gamma$ behaviors
in a similar manner to the case of shift current. 
This photo-induced Hall response is zero when the $T$ symmetry is preserved, 
since the contributions of
$\mathcal{F}$ at $\bm{k}$ and $-\bm{k}$ cancel each other in that case.   
It gives the correction to $\sigma_{xy}$ in the $T$-broken case by
the photo-excitation where $\mathcal{F}$ is the difference of the 
Berry curvatures between conduction and valence bands.
This effect is extended to the nonlinear Kerr rotation 
when the probe electric field $E(\omega=0)$ is replaced with that of
nonzero frequency.
In addition, it also expresses the effects which are finite even in the case of 
$T$-symmetric cases when one of $\bm{k}$-component is replaced by 
the parameter $Q$ characterizing the Hamiltonian. Before 
discussing this issue, let us introduce an explicit model to demonstrate the
nonlinear optical responses.

\textit{Application to inversion broken 1D chains ---}
We apply the formalism described above to a one-dimensional model described by
\begin{align}
H_0=
\sum_i& \frac{1+Q_2(-1)^i}{2} (e^{iFt} c_i^\dagger c_{i+1} + h.c.) 
\n
&+Q_1(-1)^i c_i^\dagger c_i
+ Q'_1 (-1)^i (e^{2iFt} c_i^\dagger c_{i+2} + h.c.),
\label{eq:1D}
\end{align}
where $Q_1$ is the staggered onsite energy and $Q_2$ is the bond strength alternation.
Note that the inversion symmetry is broken in the presence of {\it both} $Q_1$ and $Q_2$.
This model is describing the one-dimensional organic conductors~\cite{Su,Rice-Mele,Nagaosa-Takimoto,Onoda}.
This is also the simplest model of the ferroelectricity in perovskite materials where 
$Q_1$ corresponds to the energy level difference between the oxygen and
metal ions, and $Q_2$ to the bond strength change due to the displacement of the ions~\cite{Egami}. 
In order to see the effect of $T$-symmetry breaking, we also introduced 
$Q'_1$ which expresses complex hoppings between next nearest 
neighbors having opposite signs for two sublattices.
The Hamiltonian in Eq.(\ref{eq:1D}) is given in the $k$-space as 
\begin{align}
H_0=\cos \frac k 2 \sigma_x + Q_2\sin \frac k 2 \sigma_y 
+(Q_1 + Q'_1 \sin k)\sigma_z,
\end{align}
where $\Vec \sigma$ are the Pauli matrices describing the
degree of freedom of the two sublattices in the unit cell. 
Now let us apply an electric field $E$ to this 1D model.
The Floquet Hamiltonian is given by
\begin{align}
H_{mn}(k)&=
\begin{pmatrix}
Q_1-n\Omega & 0 \\
0 & -Q_1 -n\Omega \\
\end{pmatrix}
\delta_{mn}+
\begin{pmatrix}
C_{mn} & A_{mn}
\\
B_{mn} & -C_{mn}
\end{pmatrix},
\end{align}
with 
$A_{mn}=(t+Q_2/2)e^{-ik/2}J_{m-n}(-F/2)+(t-Q_2/2)e^{ik/2}J_{m-n}(F/2)$,
$B_{mn}=(t+Q_2/2)e^{ik/2}J_{m-n}(F/2)+(t-Q_2/2)e^{-ik/2}J_{m-n}(-F/2)$,
and $C_{mn}=Q_1'J_{m-n}(F) (-i)^{m-n}[(-1)^{m-n} e^{ik}-e^{-ik}]/(2i) $,
where $F=eEa/\Omega$ with lattice spacing $a$, and $J_n(x)$ is the $n$th Bessel function,

\begin{figure}[tb]
\begin{center}
\includegraphics[width=\linewidth]{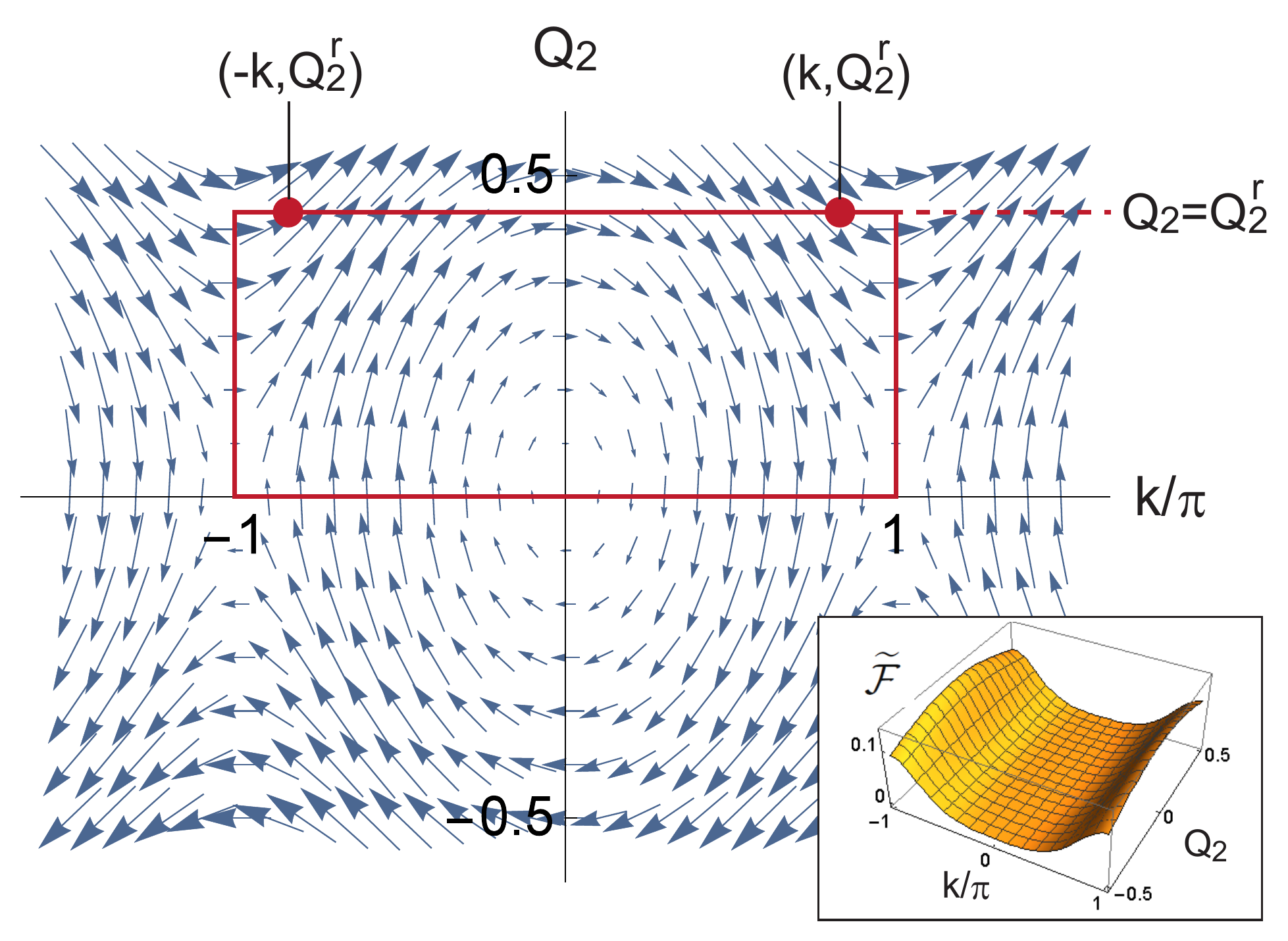}
\end{center}
\caption{
Vector field $\widetilde{\bm{\mathcal{A}}}$ for the 1D model which preserves the time-reversal 
symmetry and breaks the inversion symmetry. 
We plot 
$(\widetilde{\mathcal{A}}_k, \widetilde{\mathcal{A}}_{Q_2})$ in the parameter space $(k, Q_2)$ with $Q_1=1, Q_1'=0$.
Inset is a plot of distribution of the ``flux'' $\widetilde{\mathcal{F}}$ 
defined in the text which is related to the third order nonlinear responses.
}
\label{fig: vectorplot}
\end{figure}

We can define $R_{Q_2}$ similar to $R_k$ in Eq.~(\ref{eq:Rk})
by replacing the $k$-derivative with a $Q_2$-derivative.
Then, we can consider the two-dimensional 
vector field 
$\widetilde{\bm{\mathcal{A}}}(k,Q_2) = 
(\widetilde{\mathcal{A}}_k, \widetilde{\mathcal{A}}_{Q_2}) = 
|v^0_{12}|^2(R_k, R_{Q_2})$ in the 
$(k,Q_2)$-plane as plotted in Fig.~\ref{fig: vectorplot}.
Taking the rotation of the vector field $\bm{R}$, one can get the flux
distribution 
$\widetilde{\mathcal{F}}  =|v^0_{12}|^2( \partial_k R_{Q_2} - \partial_{Q_2} R_k )
=|v^0_{12}|^2 \mathcal{F}$
in the inset of Fig.~\ref{fig: vectorplot}.
We note that $\widetilde{\mathcal{F}}$ is 
$|v^0_{12}|^2$ times 
the difference $\mathcal{F} \equiv \mathcal{F}_2 - \mathcal{F}_1$
between the Berry curvatures of conduction and valence bands since the
contribution from the phase of the transition matrix elements drops.
These plots provide various information as follows.
First, the sum of $A_k$ at $k$ and $-k$ satisfying 
the energy conservation law $ E_2(\bm{k}) - E_1(\bm{k}) =\hbar \omega$
(indicated by two red dots in Fig.~\ref{fig: vectorplot})
corresponds to the shift-current $J$ proportional to $E^2$.  
Note that this sum does not vanish when $Q_2$ is nonzero, i.e.,
$P$ symmetry is broken.
The corresponding quantity for $\widetilde{\mathcal{A}}_{Q_2}$ gives the change in
the bond dimerization $B=\sum_i (-1)^i(c_i^\dagger c_{i+1} + h.c.)$
which is the ``current'' corresponding to the ``vector potential'' $Q_2$.
However, as one can see from Fig.~\ref{fig: vectorplot},
the contributions from $k$ and $-k$ always cancel due to the 
$T$-symmetry.

Let us now turn to the Berry curvature.
The integral of $\mathcal{F}$ over the
``first-Brillouin zone'' $-\pi < k < \pi$, $0< Q_2 < Q^r_2$ 
($Q^r_2$ is the realized value of the bond alternation),
which is denoted by a red square in Fig.~\ref{fig: vectorplot},
is related to the polarization~\cite{Resta}. 
Namely, the integral of $\mathcal{F}_1$ over the first Brillouin zone
is the polarization of the ground state. Therefore, 
that of $\mathcal{F}$ is the change of the polarization when all
the electrons in the valence band are excited to the conduction band.
The value of $\widetilde{\mathcal{F}}$ is related to the change in the
bond dimerization $B$ defined above which is proportional to 
$E^3$. 
This third order nonlinear response of $B$ is obtained if $k_y$ is replaced by $Q_2$ in   
Eqs.(\ref{eq:sigmaxy1}) and (\ref{eq:sigmaxy2}) 
and the integration over $Q_2$ is dropped.
This is intuitively understood as a ``Hall response'' of $B$ which is 
the ``current'' with respect to $Q_2$
and is transverse to the $k$-direction.
In this case,
there is no $Q_2$-integration because the contribution arises only from the realized value $Q_2^r$, and 
the photo-induced change of $B$ is given by the sum of 
$\mathcal{F}$ at $(k,Q_2^r)$ and $(-k,Q_2^r)$ with $\pm k$ satisfying 
the energy conservation law (indicated by two red dots in Fig.~\ref{fig: vectorplot}).
As shown in the inset of 
Fig.~\ref{fig: vectorplot}, the values of $\mathcal{F}$ at $k$ and $-k$ are equal to each
other, and hence this sum becomes nonvanishing.
It is useful to note here that there is a very sensitive
probe of $B$ in the case of molecular solids. The frequency shift of the 
intra-molecular vibrations detects the change of the valence state of 
each molecule~\cite{mol}.
We note that an anti-vortex in vector field $\widetilde{\bm{\mathcal{A}}}$ at $(k,Q_2)=(\pm\pi,0)$ is attributed to the peak of the Berry curvature  
while a vortex at $(k,Q_2)=(0,0)$ 
arises from the singularity in $v_{12}^0$ where $v_{12}^0$ vanishes and its phase is not well-defined.

\begin{figure}[tb]
\begin{center}
\includegraphics[width=\linewidth]{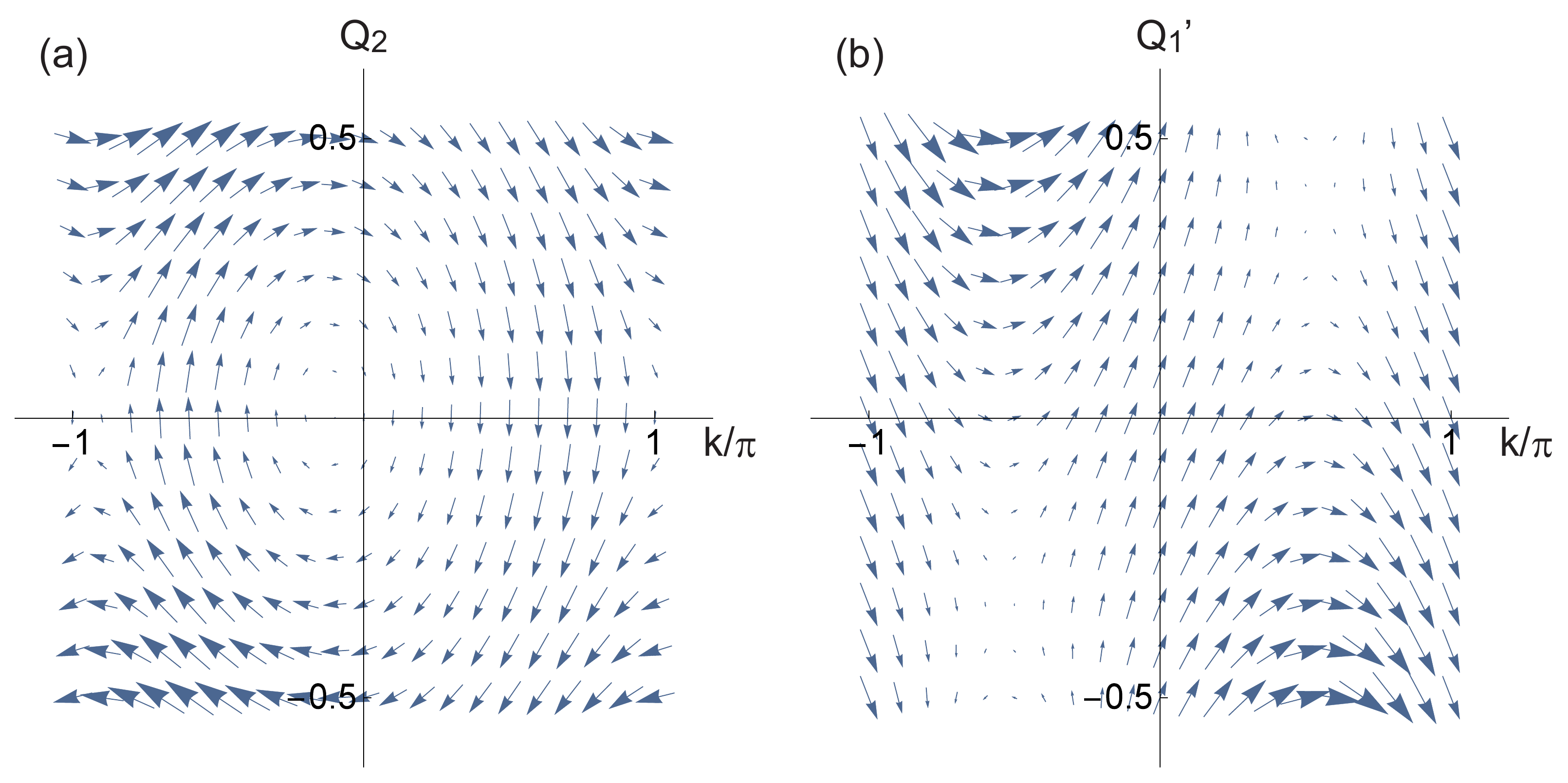}
\end{center}
\caption{
Vector fields $\widetilde{\bm{\mathcal{A}}}$ for the 1D model which break both 
time-reversal and inversion symmetries. 
We plot 
(a)$(\widetilde{\mathcal{A}}_k, \widetilde{\mathcal{A}}_{Q_2})$ with $Q_1=1, Q_1'=0.1$,
and 
(b)$(\widetilde{\mathcal{A}}_k, \widetilde{\mathcal{A}}_{Q_1'})$
with $Q_1=1, Q_2=0.4$.
}
\label{fig: vectorplot2}
\end{figure}

Next, we consider the effects of broken time-reversal symmetry $T$.
Figure~\ref{fig: vectorplot2}(a)  shows the similar plots to Fig.~\ref{fig: vectorplot}
with finite $Q'_1= 0.1$. 
It is clearly seen that the symmetry between 
$k$ and $-k$ is broken and hence all the effects discussed above can be nonvanishing.
For example, the photo-induced change in the bond dimerization 
proportional to $E^2$ becomes nonzero in addition to the shift current.
We note that there is symmetry between $Q_2$ and $-Q_2$ which originates from the $PT$ symmetry as discussed in the next section. 
Figure~\ref{fig: vectorplot2}(b) shows the vector field 
$(\widetilde{\mathcal{A}}_k, \widetilde{\mathcal{A}}_{Q_1'})$
with fixed $Q_1=1, Q_2=0.4$. 
While nonzero $Q_1$ and $Q_2$ break the $P$-symmetry,
this figure demonstrates the role of $T$-symmetry. 
Namely, both $Q_1'$ and
$k$ change their sign under $T$, and 
the vector field in Fig.~\ref{fig: vectorplot2}(b)
obeys the constraint of the $T$ symmetry.
The symmetry properties of various quantities will be discussed in the next section.

\ \par
\ \par
\textbf{Discussion:}
\ \par

\begin{figure}[tb]
\begin{center}

%
\includegraphics[width=\linewidth]{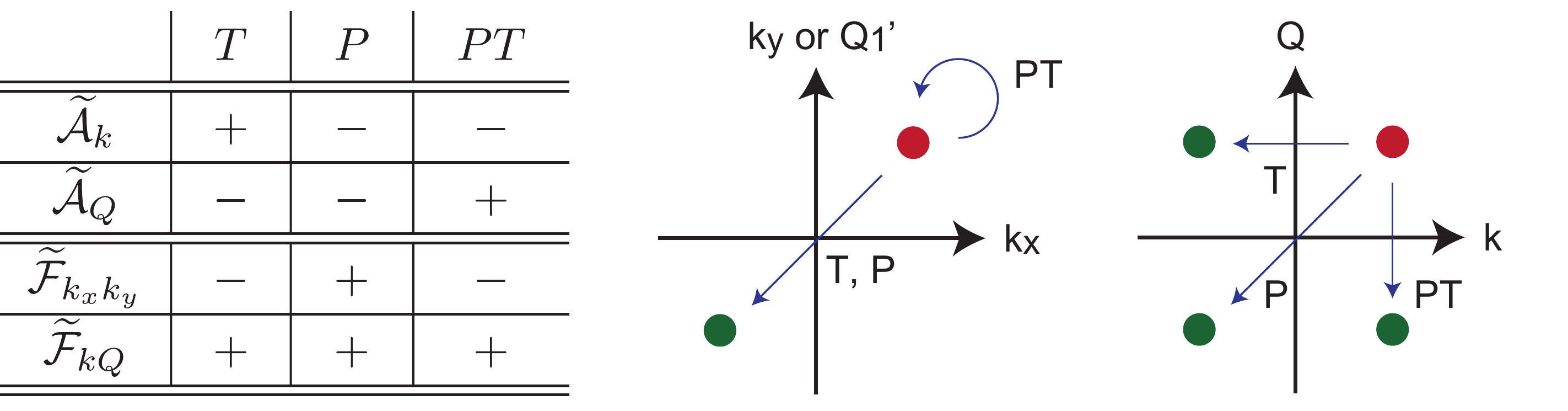}

\end{center}
\caption{
Transformation laws of Berry connection and Berry curvature.
We consider the geometry in the space spanned by the momentum $k$ 
and the parameter $Q$ quantifying the inversion breaking, and the geometry 
in the momentum space with $k_x$ and $k_y$ (or $Q_1'$).
Here, $Q$ is even under $T$, while $k$ and $Q_1'$ are odd under $T$. 
All $Q, Q_1'$ and $k$ are odd under $P$.
}
\label{fig: symmetry transformation}
\end{figure}

\textit{Symmetry considerations ---}
Based on the results presented in the previous section, it is useful to 
summarize the symmetry properties.
Figure \ref{fig: symmetry transformation} shows the transformation laws of the various quantities
with respect to $P$ and $T$. Here, the parameter $Q$ breaks 
$P$ symmetry and reverses its sign under the $P$ operation while it remains unchanged
under $T$. This is the case for $Q_2$ in the model Eq.~(\ref{eq:1D}).
On the other hand, $\bm{k}$ goes to $-\bm{k}$ for both $P$ and $T$.
The parameter $Q'_1$ is also odd under both $P$ and $T$.
The transformation properties of the Berry connection and the Berry curvature
are summarized in Fig.~\ref{fig: symmetry transformation}. 
In particular, the presence or absence of the $T$-symmetry determines whether the effect of interest is allowed or not.

Let us study these transformation properties of $\bm{\widetilde{\mathcal{A}}}$ and $\widetilde{\mathcal{F}}$ in the 1D model in Eq.~(\ref{eq:1D}) below.
First we discuss symmetry constraints on the vector fields in the $T$-symmetric case shown in Fig.~\ref{fig: vectorplot}.
The action of $T$ constrains the vector field as 
$(\widetilde{\mathcal{A}}_k (-k,Q_2),\widetilde{\mathcal{A}}_{Q_2}(-k,Q_2) )=(\widetilde{\mathcal{A}}_k (k,Q_2),-\widetilde{\mathcal{A}}_{Q_2}(k,Q_2) )$,
which is satisfied by two vectors at the two red dots in Fig.~\ref{fig: vectorplot}.
Since the nonlinear responses are contributed both from $(k,Q_2)$ and $(-k,Q_2)$,
the presence of the $T$-symmetry allows nonzero response associated with 
$\widetilde{\mathcal{A}}_k $ (shift current), 
but excludes that with $\widetilde{\mathcal{A}}_{Q_2} $ (nonlinear bond dimerization).
Similarly, the vector fields in Fig.~\ref{fig: vectorplot} 
is consistent with the constraint of the $P$-symmetry given by
$(\widetilde{\mathcal{A}}_k (-k,-Q_2),\widetilde{\mathcal{A}}_{Q_2}(-k,-Q_2) )=(-\widetilde{\mathcal{A}}_k (k,Q_2),-\widetilde{\mathcal{A}}_{Q_2}(k,Q_2) )$.
Since the flux distribution $\widetilde{\mathcal{F}}_{k Q_2}$ in the  parameter space $(k,Q_2)$ is even under both $T$ and $P$ as seen in the inset.
Since the contributions at $k$ and $-k$ always add up, nonvanishing third order nonlinear response associated with $\widetilde{\mathcal{F}}_{k Q}$ is allowed. 
(We note that the nonlinear Kerr response is not allowed by the $T$-symmetry because contributions to $\widetilde{\mathcal{F}}_{k_x k_y}$ from  $k$ and $-k$ cancel out.)
Next, we consider the cases in Fig.~\ref{fig: vectorplot2} where the $T$-symmetry is broken due to nonzero $Q_1'$. The vector field in Fig.~\ref{fig: vectorplot2}(a) is not closed under either action of $T$ or $P$ because the fixed parameter $Q_1'$ changes its sign, 
but it is closed under the combined $PT$ symmetry.
From Fig.~\ref{fig: symmetry transformation},
the vector fields are constrained by the $PT$ symmetry as 
$(\widetilde{\mathcal{A}}_k (-k,-Q_1'),\widetilde{\mathcal{A}}_{Q_1'}(-k,-Q_1') )=(\widetilde{\mathcal{A}}_k (k,Q_1'),\widetilde{\mathcal{A}}_{Q_1'}(k,Q_1') )$,
which is consistent with Fig.~\ref{fig: vectorplot2}(a).
In this case, both nonlinear responses associated with $\widetilde{\mathcal{A}}_k $ and $\widetilde{\mathcal{A}}_{Q_2}$ are allowed because the $T$-symmetry is no longer present. 
The vector field in Fig.~\ref{fig: vectorplot2}(b) is not closed under the action of $P$ because the fixed parameter $Q_2$ changes its sign under $P$,
but it is closed under the action of $T$; the $T$-symmetry  constrains the vector field in Fig.~\ref{fig: vectorplot2}(b).
Under the action of $T$, the Berry connections $\widetilde{\mathcal{A}}_k$ and $\widetilde{\mathcal{A}}_{Q_1'}$ are even as seen from Fig.~\ref{fig: symmetry transformation}.
Thus the vector fields transform as
$(\widetilde{\mathcal{A}}_k (-k,-Q_1'),\widetilde{\mathcal{A}}_{Q_1'}(-k,-Q_1') )=(\widetilde{\mathcal{A}}_k (k,Q_1'),\widetilde{\mathcal{A}}_{Q_1'}(k,Q_1') )$
which is consistent with Fig.~\ref{fig: vectorplot2}(b).

\textit{Conclusions ---}
We have studied the nonlinear optical responses 
from the topological properties based on the Keldysh + Floquet formalism 
taking into account the two states connected by the optical transition.
The Berry connection and the Berry curvature appear in the even order and odd order
responses in the electric field $E$ of the light, respectively.
For example, the shift-current proportional to $E^2$
is represented by the Berry connection, while
the Berry curvature appears in the third order response in $E$. 
These processes involve the excitation of electrons from the valence band to the conduction band, 
where created electrons and holes have been assumed to be non-interacting in this paper.
In real materials, however, the electron correlation effect should be taken into account.
In particular, the excitonic effect will hinder the photo-current generation.
Therefore the many-body formulation of the 
nonlinear optical responses is an important issue to be studied in the future. 
As for the ferroelectric materials, however, the large dielectric constant screens the
Coulomb effect and the excitonic effect is suppressed which may justify
the single-particle treatment.

\textit{Acknowledgment ---}
We thank Y. Tokura, M. Kawasaki, N. Ogawa, J.~E. Moore, J.~Orenstein and B.~M. Fregoso for fruitful discussions.
 This work was supported by 
the EPiQS initiative of the Gordon and Betty Moore Foundation (TM), and by JSPS
Grant-in-Aid for Scientific Research
(No. 24224009, and No. 26103006) from MEXT, Japan, 
and ImPACT (Impulsing Paradigm Change through Disruptive Technologies) Program of Council for Science, Technology and Innovation (Cabinet office, Government of Japan) (NN).

\begin{widetext}
\ \par
\ \par
\textbf{Method:}
\ \par

\textit{Keldysh Green's function---}

The Keldysh Green's function in the Floquet formalism is given by
the Dyson equation~\cite{Jauho,Kamenev,Kohler,Oka},
\begin{align}
\begin{pmatrix}
G^R & G^K \\
0 & G^A
\end{pmatrix}^{-1}_{mn}
&=
\begin{pmatrix}
(\omega+n\Omega)\delta_{mn}-H_{mn} & 0 \\
0 & (\omega+n\Omega)\delta_{mn}-H_{mn}
\end{pmatrix}
+\Sigma_{mn},
\end{align}
where $m,n$ run over the Floquet indices, and $\Sigma$ is the self energy.
The Floquet Hamiltonian $H$ is obtained by expanding a Hamiltonian $H(t)$ periodic 
in time with period $T$ in the Floquet modes as
\begin{align}
H_{mn}&=
\frac{1}{T}\int_0^T dt e^{i(m-n)\Omega t} H_0(t),
\label{eq: floquet modes}
\end{align}
with $\Omega=2\pi/T$.
We assume that each site is coupled to a heat reservoir with the Fermi distribution 
function $f(\epsilon)$ with a coupling constant $\Gamma$.
In this case, the self energy is written as
\begin{align}
\Sigma_{mn}&=
i\Gamma \delta_{mn}
\begin{pmatrix}
\frac 1 2 & -1+2f(\omega+m\Omega) \\
0 & -\frac 1 2
\end{pmatrix}.
\label{eq: Sigma}
\end{align}
Then the current is given by
\begin{align}
J(t)&=\sum_{m}-i\t{Tr}[v(t) G^<_{mn}] e^{-i(m-n)\Omega t},
\label{eq: J(t)}
\end{align}
where $v(t)$ is the time-dependent velocity operator defined by
$v(t)=\partial H_0(t)/\partial k$
and $\t{Tr}$ denotes an integration over $k$ and $\omega$ and a trace over band indices (but not over Floquet indices).
We note that the reference Floquet index $n$ can be arbitrarily chosen due to the translation symmetry in the Floquet index.
The lesser Green's function $G^<$ is given by
\begin{align}
G^<&=G^R \Sigma^< G^A, \\
\Sigma^< &=\frac{\Sigma^R+\Sigma^K-\Sigma^A}{2} .
\label{eq: G<}
\end{align}
We note that the retarded and advanced Green's functions are simply written as
\begin{align}
(G^{R/A})^{-1}_{mn}
&=
\left(\omega+n\Omega\pm \frac{i\Gamma}{2} \right) \delta_{mn}
-H_{mn}.
\end{align}
Furthermore, the dc part of the current $J$
 is concisely obtained from the dc current operator defined from the Floquet Hamiltonian as
\begin{align}
J&=\sum_{m}-i\t{Tr}[\tilde{v}_{nm} G^<_{mn}], \\
\tilde{v} &= \frac{\partial H_F}{\partial k}.
\end{align}


\textit{Lesser Green's function for the Floquet two band model---}
In this section, we focus on the Floquet two band model and study the lesser Green's function that is directly related to physical quantities.

First, we derive the Floquet Hamiltonian $H_F$ starting from the original Hamiltonian without a drive $H_\t{orig}(k)$.
In the presence of the monochromatic light $E(t)=E e^{-i\Omega t}+E^* e^{i\Omega t}$,
the time-dependent Hamiltonian is given by
\begin{align}
H_0(t)&=H_\t{orig}(k+A(t)), \\
A(t)&=i A e^{-i\Omega t} -i A^* e^{i\Omega t},
\end{align}
with $A=E/\Omega$.
By keeping terms up to the linear order in $A$, one obtains 
\begin{align}
H_0(t)&\cong H_\t{orig}(k) + A(t) v^0,
\end{align} 
with $v^0=\partial H_\t{orig}/\partial k$.
Next we express this time-dependent Hamiltonian as a Floquet Hamiltonian $(H_F)_{mn}=H_{mn} -n\Omega \delta_{mn}$ by using Eq.~(\ref{eq: floquet modes}).
We further focus on two Floquet bands, i.e., the valence band with the Floquet index $n=-1$ and the conduction band with the Floquet index $n=0$. 
This leads to the two by two Floquet Hamiltonian
\begin{align}
H_F&=
\begin{pmatrix}
\epsilon^0_1 +\Omega & -i A^* v^0_{12} \\
iA v^0_{21} & \epsilon^0_2
\end{pmatrix}
\equiv
\epsilon+\Vec d \cdot \Vec \sigma,
\end{align}
where
the subscripts 1 and 2 refer to the valence band and conduction band, respectively,
and $\epsilon^0_i=(H_\t{orig})_{ii}$.

For this two by two Floquet Hamiltonian, the lesser Green's function $G^<$ is obtained as follows.
We consider the case where a coupling to a heat bath leads to the self energy given by Eq.~(\ref{eq: Sigma}).
In this case, the retarded and advanced Green's function are written as
\begin{align}
G^{R/A}&=
\frac{\omega-\epsilon \pm i\Gamma/2+\bm d \cdot \bm \sigma}
{(\omega-\epsilon \pm i\Gamma/2)^2-d^2}.
\end{align}
Since the Fermi energy of the bath lies within the energy gap of the system, the Keldysh component of the self energy reduces to
\begin{align}
\Sigma^<&=i\Gamma \frac{1+\sigma_z}{2}.
\end{align}
With these data, the lesser Green's function for the Floquet two band model is obtained from Eq.~(\ref{eq: G<}).
For example, an off-diagonal element of $G^<$
is given by
\begin{align}
(G^<)_{21}&=
\frac{(d_x + id_y) \left(\frac{\Gamma}{2} + id_z \right)}{2\left(d^2+\frac{\Gamma^2}{4} \right)}.
\label{eq: G< 21}
\end{align}
Here we note that the superscripts 21 indicate bases of the two by two Hamiltonian $H_F$ 
(with corresponding Floquet indices $0,-1$ ), 
and $(G^<)_{21}$ describes the Fourier component of $e^{-i\Omega t}$
according to Eq.~(\ref{eq: J(t)}). 
Moreover,
general expectation values are written as
\begin{align}
\braket{b_0+\bm b \cdot \bm \sigma}
&=
-i\t{Tr}[(b_0+\bm b \cdot \bm \sigma) G^<] \n
&=
\frac{1}{d^2+\frac{\Gamma^2}{4}}
\left[ 
\frac{\Gamma}{2}(-d_x b_y+ d_y b_x)
+
(d_x b_x + d_y b_y) d_z
+
\left(d_z^2+\frac{\Gamma^2}{4} \right) b_z
\right] 
+b_0.
\label{eq: general expectation value}
\end{align}

\end{widetext}

\bibliography{photovoltaics.bib}

\end{document}